# Ultra-low Hysteresis in Giant Magnetocaloric $Mn_{1-x}V_xFe(P, Si, B)$ Compounds

Jiawei Lai[1,2*], Bowei Huang[1,2], Dimitrios Bessas,[2] Xinmin You[2], Michael Maschek[2],

Dechang Zeng[1*], L. Zhang[2], Niels van Dijk[2], Ekkes Brück[2]

1. *School of Materials Science & Engineering, South China University of Technology, Guangzhou 510640, China;*

2. *Fundamental Aspects of Materials and Energy, Department of Radiation Science and Technology, TU Delft, Mekelweg 15, 2629JB Delft, The Netherlands*

## Abstract

Large thermal hysteresis in the MnFe(P, Si, B) system hinders the heat exchange rate and thus limits the magnetocaloric applications at high frequencies. Substitution of Mn by V in $Mn_{1-x}V_xFe_{0.95}P_{0.593}Si_{0.33}B_{0.077}$ and $Mn_{1-x}V_xFe_{0.95}P_{0.563}Si_{0.36}B_{0.077}$ alloys was found to reduce the thermal hysteresis due to a decrease in the latent heat. Introducing V increases both the field-induced transition temperature shift and the magnetic moment per formula unit. Thus, a decease in the thermal hysteresis is obtained without losing the giant magnetocaloric effect. In consequence, an ultralow hysteresis (0.7 K) and a giant adiabatic temperature change of 2.3 K were achieved, which makes these alloys promising candidates for commercial magnetic refrigerator using permanent magnets.

# 1. Introduction

Room temperature magnetic refrigeration technology has attracted broad attention due to its advantages compared to the traditional cooling techniques, such as a high efficiency and a low impact on the environment. [1] Materials with a giant magnetocaloric effect (GMCE), which are the working refrigerant of this technique, will release heat when an external magnetic field is applied, while they absorb heat when the magnetic field is removed. The performance of GMCE materials mainly determine the working efficiency and the cooling power of this technology. The GMCE usually occurs in materials that show a first-order magnetic transition (FOMT), such as $Gd_5Ge_2Si_2$,[2] $LaFe_{13-x}Si_x$,[3,4,5] $MnFeP_{1-x}As_x$, [6] $MnFeP_{1-x-y}Si_xB_y$, [7,8,9] $MnCoGeB_x$[10] and Heusler [11] alloys. Among them, the $MnFeP_{1-x-y}Si_xB_y$ alloys are regarded as one of the most promising materials that can be industrialized as magnetic refrigerant because of their cheap and non-toxic elements, high cooling capacity and tunable $T_C$ near room temperature. [7]

Thermal hysteresis ($\Delta T_{hys}$) is an important issue that limits the real application of the GMCE in these FOMT materials. [12] The discontinuous nature of the transition is the feature that provides the GMCE. Therefore, in the premise of keeping the GMCE, the thermal hysteresis should be made as narrower as possible by manipulating the microstructure or by tuning the composition. Through 0.075 at.% of B substitution in the $MnFeP_{1-x-y}Si_xB_y$ alloys, the optimized $\Delta T_{hys}$ can be decreased to 1.6 K according to temperature-dependent magnetization curves at a magnetic field of 1 T and $\Delta T_{hys}$ is 2.0 K according to in-field DSC measurements at a magnetic field of 1 T (see the supporting information of ref [9]), while maintaining a GMCE. [9] In this case, the material can be cycled for 10 thousand times and the sample geometry remains intact. A higher level of B substitution can decrease the $\Delta T_{hys}$ further, but fail to provide a sufficiently large GMCE. [13] It is essential to find a new approach to further decrease the $\Delta T_{hys}$ and simultaneously provide a large GMCE. One of the design criteria is that the adiabatic temperature change ($\Delta T_{ad}$) should be larger than 2 K[14], since Engelbrecht and Bahl [15] demonstrated that cooling may be ineffective when $\Delta T_{ad}$ drops below 2 K. In this work, through V substitution, an ultra-low $\Delta T_{hys}$ (0.7 K) and a GMEC of $\Delta T_{ad}$ (2.3 K) at a magnetic field of 1 T is achieved simultaneously.

The crystal structure of $MnFeP_{1-x-y}Si_xB_y$ shows a significant change in lattice parameters across the magnetic phase transition, while it keeps its hexagonal structure (magneto-elastic transition).[16,9] Applying a magnetic field results in a shift of the transition temperature ($T_c$) to higher temperatures. The shift of $T_c$ induced by magnetic fields, defined as $dT_c/dB$, is positive for a conventional first-order magnetic transition materials such as $MnFeP_{1-x-y}Si_xB_y$ [13] and La-Fe-Si[17], while it is negative for the inverse first-order magnetic transition materials, for instance the Ni–Mn–X–Heusler alloys with X = Sn, Sb and In [18] or Fe-Rh[19]. For the conventional first-order magnetic transition materials, this shift is attributed to the magnetic field stabilization of the phase with the higher magnetization, being the low-temperature ferromagnetic phase [20,21]. In a magnetic field thermal energy is then needed to induce the magnetic phase transition. If the value of $dT_c/dB$ is enhanced, the magnetic phase transition can be induced in lower magnetic field. As a consequence, low-field permanent magnets could be utilized, which would significantly reduce the costs of commercial applications. The magnetic field currently used in the commercial prototypes is generated by NdFeB permanent magnets with external magnetic fields varying from an 0.9 to 1.5 T [22,23,24]. The materials cost to reach a field of 1.5 T may be 10 times higher than the costs to reach a field of 0.9 T. It is therefore of interest to explore the lower field potential of this GMCE system by studying $dT_c/dB$. In this work, we investigated the effect of V substitution on the $\Delta T_{hys}$, $dT_c/dB$, the lattice parameters and the magnetic properties in polycrystalline Mn-V-Fe-P-Si-B alloys.

## 2. Experimental

Polycrystalline $Mn_{1-x}V_xFe_{0.95}P_{0.593}Si_{0.33}B_{0.077}$ ($x = 0.00, 0.01, 0.02, 0.03$) alloys were prepared by a powder metallurgy method. The starting materials in the form of Mn, Fe, red P, Si and V powders were mechanically ball milled for 10 h in an Ar atmosphere with a constant rotation speed of 380 rpm, then pressed into small tablets, and finally sealed in quartz ampoules under 200 mbar of Ar before employing the various heat treatment conditions. These tablets were annealed at 1323 K for 2 h in order to crystalize and slowly cooled down to room temperature. Then they were heated up to the same annealing temperature for 20 h to homogenize the alloy and finally quenched in water. This batch samples is regarded as *series A*. In order to tune the $T_C$ to room temperature for the sample with V, the $Mn_{1-x}V_xFe_{0.95}P_{0.563}Si_{0.36}B_{0.077}$ ($x = 0.00, 0.01, 0.02, 0.03$) alloys with a higher Si content were prepared with the same procedure as series *A*, except for a higher annealing temperature of 1373 K. This batch samples is regarded as *series B*.

The X-ray diffraction (XRD) patterns were collected on a PANalytical X-pert Pro diffractometer with Cu-Kα radiation (1.54056 Å) at room temperature. The temperature and magnetic field dependence of the magnetization was measured with a commercial superconducting quantum interference device (SQUID) magnetometer (Quantum Design MPMS 5XL) in the reciprocating sample option (RSO) mode. The adiabatic temperature change ($\Delta T_{ad}$) is measured in a Peltier cell based differential scanning calorimetry using a Halbach cylinder providing a magnetic field of 1.5 T. In this setup, the iso-field calorimetric scans were performed at a slow rate of 50 mKmin$^{-1}$ in order to probe the equilibrium state, while the temperature has been corrected for the effect of the thermal resistance of the Peltier cells.

## 3. Results and discussions

In figure 1 the XRD patterns for *series A* (Fig. 1*a* and 1*b*) and *series B* (Fig. 1*c* and 1*d*) are illustrated. For the $Mn_{1-x}V_xFe_{0.95}P_{0.563}Si_{0.36}B_{0.077}$ ($x = 0.00$, 0.01) alloys in *series B* , as $T_C$ is higher than room temperature, the XRD patterns are measured at 323 K, where they are in the paramagnetic state. Other samples are measure at room temperature since their $T_C$ values are below room temperature. At the selected temperatures, all the samples are measured at paramagnetic state. The hexagonal Fe$_2$P-type (space group P-*62m*) phase is identified as the main phase in all these alloys and the cubic MnFe$_2$Si-type phase (space group F*m3m*) is identified as the impurity phase. Based on the refinement results, the estimated fraction of the impurity phase is 1.6 - 2.4 vol.% in *series A* and 3.7 - 4.5 vol.% in the *series B*, respectively. The amount of impurity phase is decreases by V substitution for *series B*. The lattice parameter in *series A* and *series B* shows a different behavior. For *series A*, an increase in V substitution leads to a decrease in the *a* axis and an increase in the *c* axis. The *c/a* ratio increases, while $T_C$ decreases for an increasing V substitution. Note that, the unit cell volume of the crystal remains unchanged for $x = 0.01$ and 0.02. Only when the V content reaches $x = 0.03$, the volume drops by 0.7% compared to $x = 0.00$. For *series B*, the lattice parameters show a different trend. Oppositely, an increasing V substitution leads to an increase in the *a* axis, while the *c* axis decreases for $x = 0.00$, 0.01 and 0.02. The evolution of the unit cell volume  for *series B* was found to differ from *series A* as the unit cell volume slightly increases for $x = 0.02$ and 0.03, but still smaller than $x = 0.00$. Since the covalent radius of V ($132 \pm 5$ pm) is slightly smaller than that of Mn ($139 \pm 5$ pm), a decrease in the unit cell volume may be a sign of the substitution of Mn by V in the $Fe_2P$-type structure.

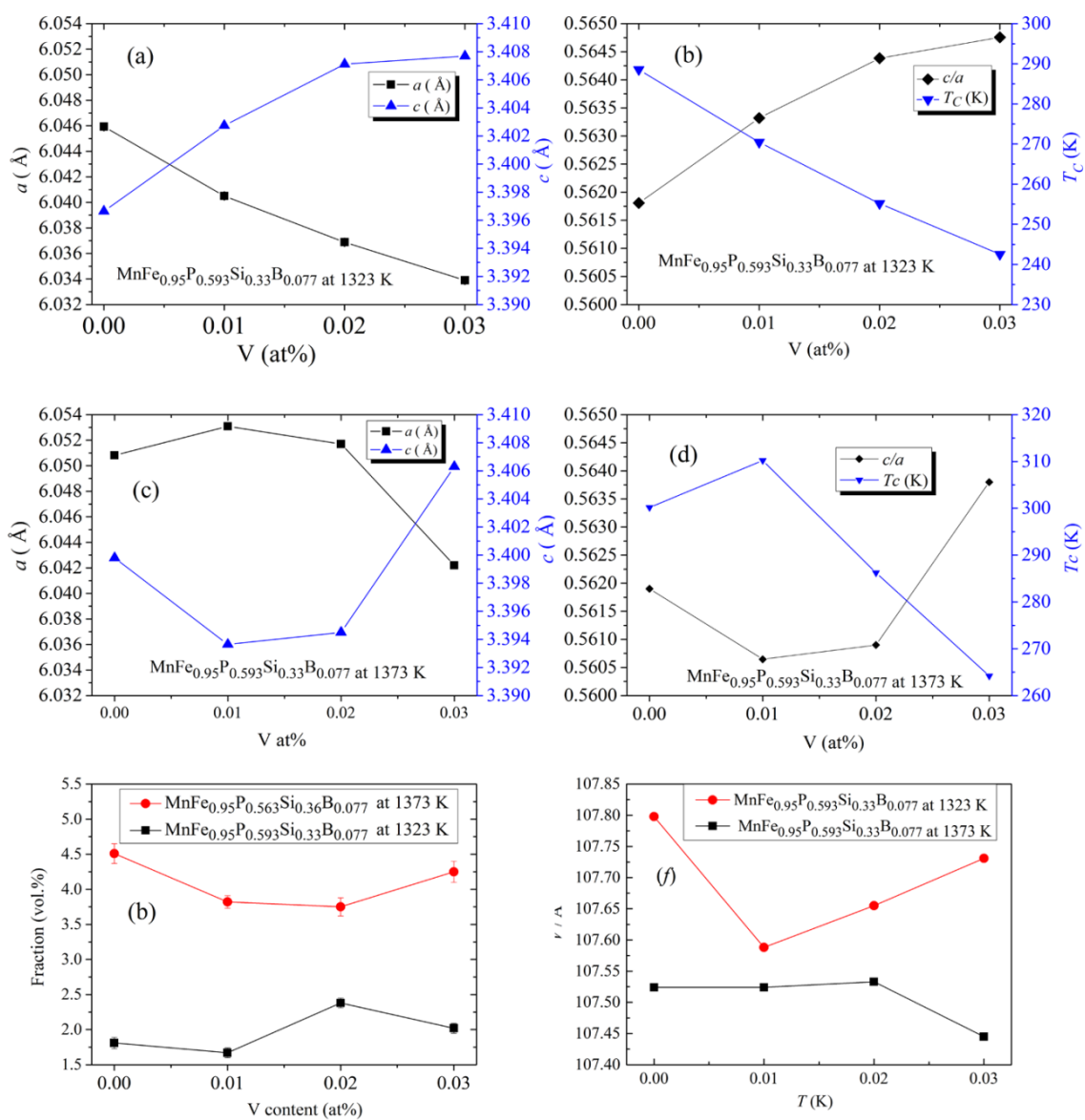

Fig. 1 (*a*) Lattice parameters of the *a* and *c* axis for *series A* and (*b*) *c/a* ratio and $T_C$ for series A; (*c*) Lattice parameters of the *a* and *c* axis for *series B* and (*d*) *c/a* ratio and $T_C$ for *series B* ; (*e*) Fraction of the second phase in *series A* and *B* ; (*f*) Unit cell volume of *series A* and *B*.

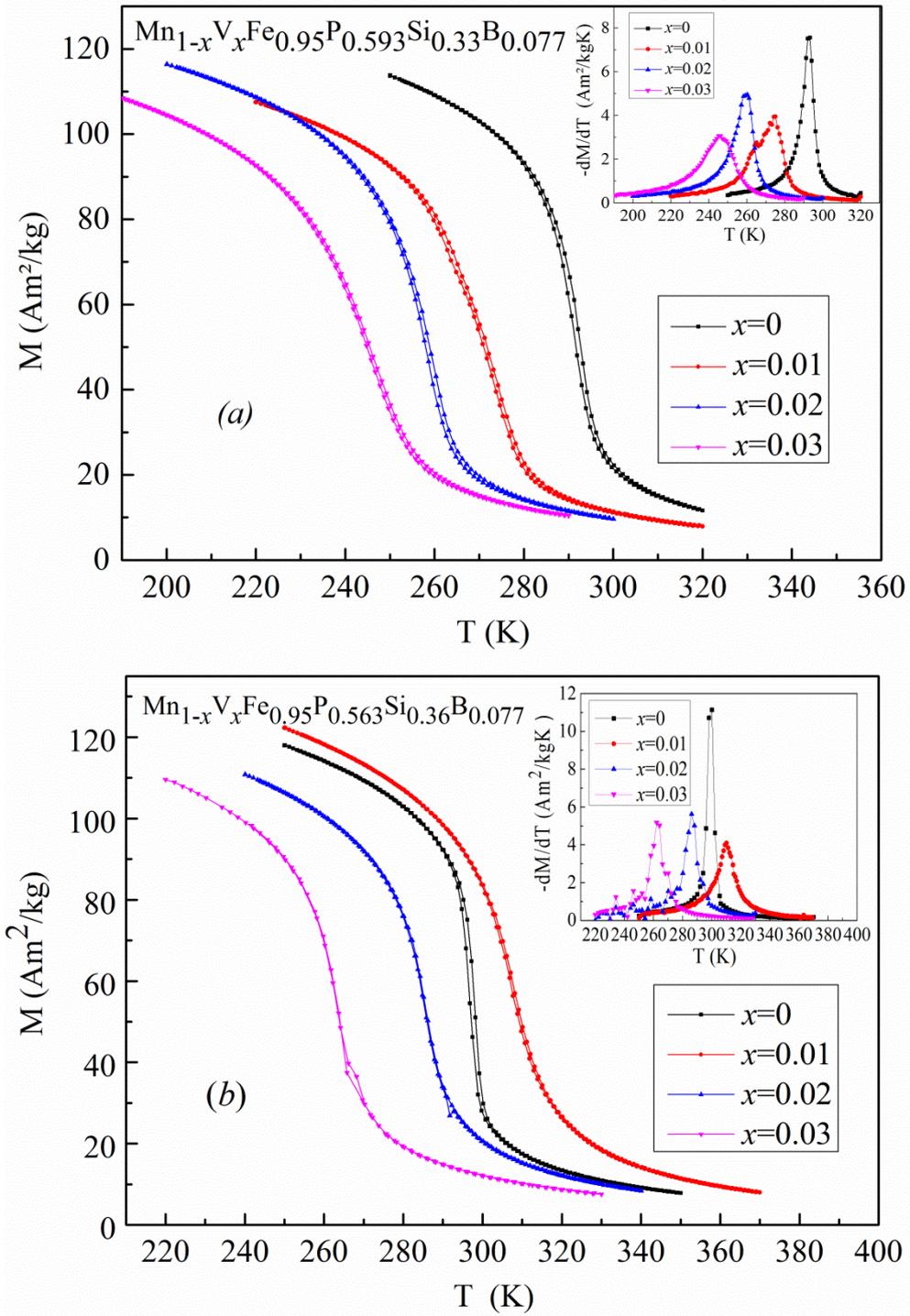

Fig. 2 (a) Temperature dependence of the magnetization in *series A* under an applied magnetic field of 1 T; (b) Temperature dependence of the magnetization in *series B* under an applied magnetic field of 1 T.

Table 1 The Curie temperature ($T_C$), thermal hysteresis ($\Delta T_{hys\text{-}MT}$), latent heat ($L$), magnetic entropy change ($|\Delta S_M|$) and adiabatic temperature change $\Delta T_{ad}$ at a field change of 1 T for *series A* and *B*.

| Annealed T (K) | Sample | $T_C$ (K) | $\Delta T_{hys\text{-}MT}$ (K) | $L$ (kJ/kg) | $|\Delta S_M|$ (J/(kg·K)) | $\Delta T_{ad}$ (K) |
|---|---|---|---|---|---|---|
| *series A* | $x = 0.00$ | 290.0 | 1.1 | 5.2 | 6.5 | 2.7 |
| *series A* | $x = 0.01$ | 270.4 | 0.8 | 3.4 | 3.3 | / |
| *series A* | $x = 0.02$ | 255.2 | 0.9 | 2.4 | 4.6 | 1.6 |
| *series A* | $x = 0.03$ | 242.5 | 0.7 | 1.7 | 2.7 | / |
| *series B* | $x = 0.00$ | 300.2 | 1.5 | 6.2 | 11.3 | 3.5 |
| *series B* | $x = 0.01$ | 310.2 | 0.8 | 2.5 | 4.8 | 1.8 |
| *series B* | $x = 0.02$ | 286.2 | 0.5 | 3.7 | 5.6 | 2.3 |
| *series B* | $x = 0.03$ | 264.2 | 0.1 | 2.8 | 4.8 | 1.6 |
| Ref. [9]* | $x = 0.00$ | 281 | 1.6 | 3.8 | 9.8 | 2.5 |

Temperature dependence of the magnetization in *series A* and *B* are shown in figure 2 (a) and (b), respectively. The temperature dependence of *-dM/dT* is also shown in the corresponding insets. Generally, the maximum of *-dM/dT* is regarded an indication of the strength for FOMT. The maximum of *-dM/dT* in our materials decreases for an increasing V content except for the sample with $x = 0.02$, indicating it moves closer to SOMT. The reason why the sample with $x = 0.02$ has this jump is unclear. The transition temperature $T_C$ is determined from the maximum value of the *-dM/dT* in the *M-T* curve during heating. For *series A*, $T_C$ tends to decrease with increasing V substitution. Moreover, the reduction of $T_C$ becomes weaker with increasing V content, as shown in table 1. It reduces from about 18.1, 15.3 and 12.7 K from $x = 0.00$ to 0.03 in steps of 0.01 at.% V. For *series B* , $T_C$ first increases at $x = 0.01$ and then decreases with increasing V substitution.

The DSC patterns for *series A* and *B* are measured (not shown here), and the derived latent heat is listed in Table 1.  In the previous, $Mn_1Fe_{0.95}P_{0.593-x}Si_{0.33}B_x$ alloys annealed at 1373 K in two-step heat treatment were studied [9]. Note that, it was found that the alloy corresponding to the composition with  $x = 0.00$ in this work is already at the border of the FOMT to the second order magnetic transition (SOMT). Increasing the V substitution from 0.00 to 0.03 results in a strong reduction of the latent heat by 67% from 5.2 to 1.7 J/g  for the alloys annealed at 1323 K and by 55% from 6.2 to 2.8 J/g for the alloys annealed at 1373 K (listed in table 1), indicating that the samples transfer more towards the SOMT. As mentioned above, the reduction in latent heat mainly contribute to the increase in d$T_C$/d$B$. A smaller latent heat will result in a smaller thermal hysteresis.

A large $\Delta T_{hys}$ is usually accompanied with a strong FOMT in the materials families of $Gd_5(Si,Ge)_4$[25,26], $La(Fe,Si)_{13}$[21], and Heuslers $NiMn(In,Ga,Sn)$[27] and $(Mn,Fe)_2(P,Si,B)$ [28] alloys. Even though they have a giant MCE, the large $\Delta T_{hys}$ limits their application in real devices since it will lower the heat exchanging efficiency dramatically. Materials optimized to be near the critical point between a first and second order transition are promising candidates for applications as they combine a low thermal hysteresis with a considerable GMCE [9]. Here, we find that $\Delta T_{hys}$ can be reduced further by substituting Mn by V in $(Mn,Fe)_2(P,Si,B)$ alloys. $\Delta T_{hys-MT}$ is determined by calculating the difference in the maximum value of -$dM/dT$ during cooling and heating in an applied magnetic field of $\mu_0H = 1$ T. For *series A*, $\Delta T_{hys-MT}$ decreases by 36% from 1.1 to 0.7 K when $x$ increases from 0.00 to 0.03.

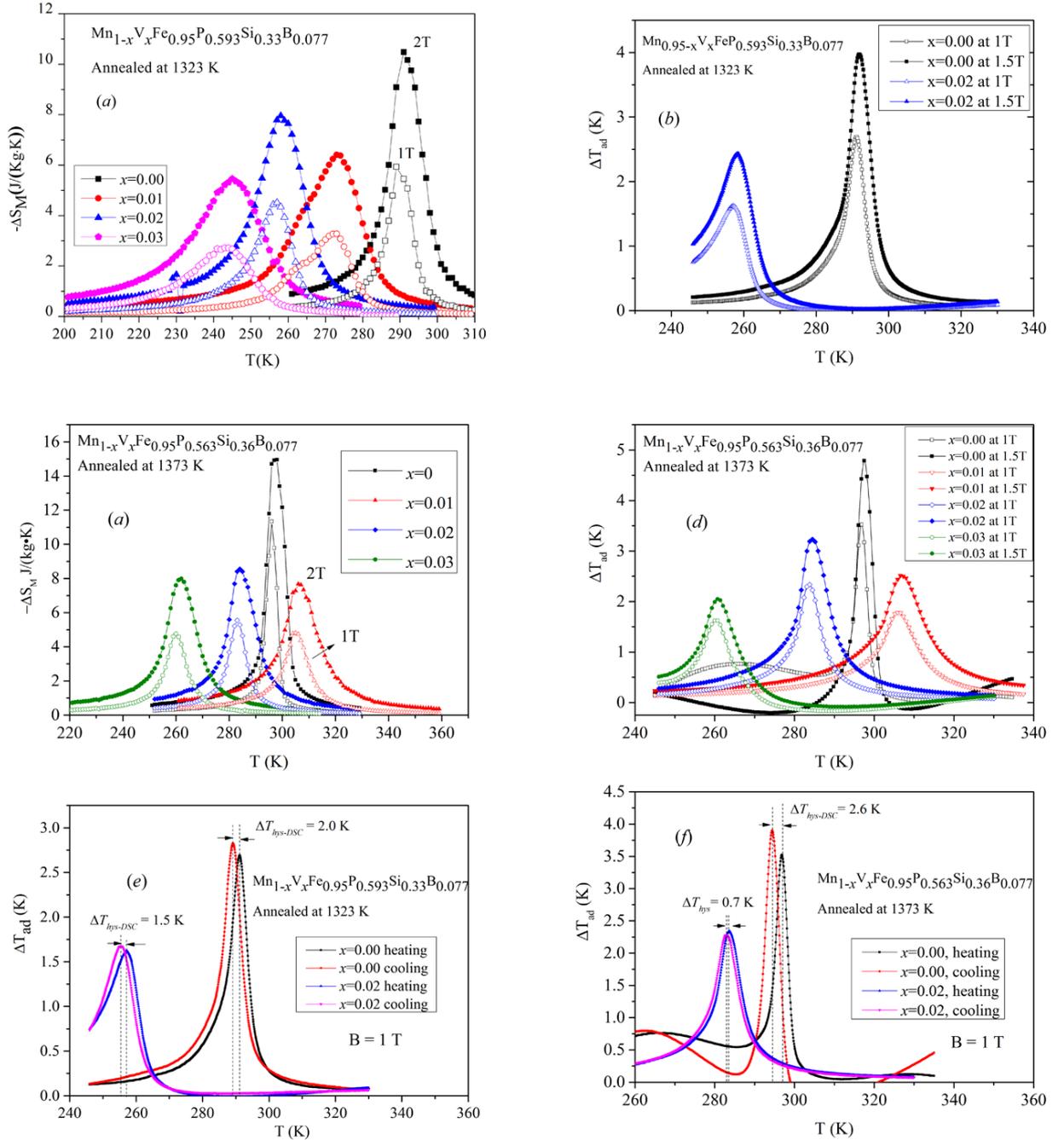

Fig. 3 (*a*) and (*c*) Temperature dependence of |$\Delta S_M$| under a field change of 0-1.0 T (open symbols) and 0-2.0 T (filled symbols) for *series A* and *B*, respectively; (*b*) and (*d*) Temperature dependence of $\Delta T_{ad}$ under a field change of 0-1.0 T (open symbols) and 0-1.5 T (filled symbols) for *series A* and *B*, respectively; (*e*) Partial temperature dependence of $\Delta T_{ad}$ under a field change of 0-1.0 T during heating and cooling for *series A*; (*f*) Partial temperature dependence of $\Delta T_{ad}$ under a field change of 0-1.0 T during heating and cooling for *series B* .

For *series B* , $\Delta T_{hys\text{-}MT}$ decreases by 93% from 1.5 to 0.1 K when *x* increases from 0.00 to 0.03. The thermal hysteresis decreases with increasing V substitution, which tunes the *series A* and *B* alloys towards a second order magnetic transition which makes these materials more suitable for commercialization of magnetic refrigerators.

The iso-field magnetization curves (not shown here) of *series A* and *B* for a magnetic field change of 0-2 T are measured in the vicinity of $T_C$ with a temperature interval of 1 K. The values of $|\Delta S_M|$ for the alloys is derived from extracted isothermal magnetization curves using the Maxwell relation [29, 30]. The temperature dependence of $|\Delta S_M|$ for *series A* and *B* are shown in figure 3 (*a*) and (*c*), respectively. $|\Delta S_M|$ decreases with increasing V substitution. However, the alloy with *x* = 0.02 in *series A* has a higher $|\Delta S_M|$ value even though it has a lower latent heat. In *series A*, the MCE ($|\Delta S_M|$ = 6.5 J/(kgK) at 289 K under a field change of 0-1 T with $\Delta T_{hys}$ = 1.1 K) of the alloy with *x* = 0.00 is comparable to the previously studied one [9] prepared by a second step annealing method ($|\Delta S_M|$ = 9.2 J/(kgK) at 279.1 K under a field change of 0-1 T with $\Delta T_{hys}$ = 1.6 K).

Figure 3 (*b*) illustrates the temperature dependence of in-field DSC values of $\Delta T_{ad}$ for a partial *series A* (*x* = 0.00 and 0.02), while figure 3 (*d*) illustrates the temperature dependence of $\Delta T_{ad}$ for *series A* (*x* = 0.00, 0.01, 0.02 and 0.03). When *x* increases from 0.00 to 0.02 in *series A*, the value of $\Delta T_{ad}$ decreases from 2.7 to 1.6 K under a field change of 1 T. When *x* increases from 0.00 to 0.02 in *series B* , the values of $\Delta T_{ad}$ decreases from 3.5 to 2.3 K under a field change of 1 T. Note that, in *series B*, the value of $\Delta T_{hys\text{-}DSC}$, determined by the difference of the heating and cooling process of in-field DSC under a field change of 1 T, decreases from 2.4 to 0.7 K when *x* increases from 0.00 to 0.02. The value of $\Delta T_{ad}$ for $Mn_{0.98}V_{0.02}Fe_{0.95}P_{0.563}Si_{0.36}B_{0.077}$ ($\Delta T_{ad}$ = 2.3 K) in *series B* is competitive to the MnFe$_{0.95}$P$_{0.563}$Si$_{0.36}$B$_{0.077}$ alloys ($\Delta T_{ad}$ = 2.5 K) [9] , but its value of $\Delta T_{hys-DSC}$ is reduced by 85%. In addition, for the *series B*, the latent heat calculated from the calorimetry measurement (which is not shown here) is $5.1 \pm 0.02$ and $3.3 \pm 0.02$ J/g for $x$=0.00 and $x$= 0.02, respectively. It is clearly promising to achieve at the same time a giant value of $\Delta T_{ad}$ and an extremely low $\Delta T_{hys-DSC}$, which can significantly improve the heat exchange efficiency of the magnetic cooling system.

The magnetic field dependence of $T_C$ and d$T_C$/d$B$ for *series A* and *B* are shown in figure 4 (a) and (b). The magnetic field (on the horizontal axis) has been corrected by the demagnetizing field using a demagnetization factor of 1/3, as the shape of measuring powders can be simplified as spheres. In order to demonstrate the change in d$T_C$/d$B$, the value of $T_C$ (B)-$T_C$ (0) versus the magnetic field is shown in figure 4 (*a*) and (*b*). The Clausius–Clapeyron relation for a FOMT corresponds to $dT_C/dB = -T_C \Delta M/L$, where $B$ is the applied magnetic field and $\Delta M$ is the jump in magnetization, implying that d$T_C$/d$B$ should increase with an increase of $\Delta M$ and a decrease of the latent heat [**Error! Bookmark not defined.**]. For the alloys annealed at 1323 and 1373 K, d$T_C$/d$B$ can be enhanced from 4.0 to 5.0 K/T when the V content is changed from $x = 0.00$ to $x = 0.02$. A value of 5.0 K/T is comparable to the d$T_C$/d$B$ value of (Mn,Fe)$_2$(P,As) alloys, where d$T_C$/d$B$ was found to be 5.2 K/T. [31] This increase is mainly caused by the decrease of the latent heat (see table 1) since the values of $T_C$ and $\Delta M$ are reduced (see Figure 3). Moreover, figure 4 (c) demonstrates that the magnetic moment per formula unit ($\mu_{f.u.}$) for *series B* increases from 3.75 to 3.96 $\mu_B/_{f.u.}$ when $x$ increases from 0.00 to 0.02. The value of $\mu_{f.u.}$ for *series B* was calculated as mentioned in reference [32]. A larger value of $\mu_{f.u.}$ suggests a larger value of $/\Delta S_M/$. The higher values for d$T_C$/d$B$ and $\mu_{f.u.}$ explains why a ultra-low thermal hysteresis and a giant GMEC can be achieved simultaneously in the alloys with V. By B substitution, the thermal hysteresis reaches a minimum, while $\Delta T_{ad}$ remains 2 K.

Introducing V as a new substitutional element is found to be capable of increasing both $dT_C/dB$ and $\mu_{f.u.}$ and can further decease the hysteresis without losing the GMCE. Thus, the current $Mn_{1-x}V_xFe(P,Si,B)$ compounds provide a feasible alternative for high-frequency near room temperature magnetic cooling applications.

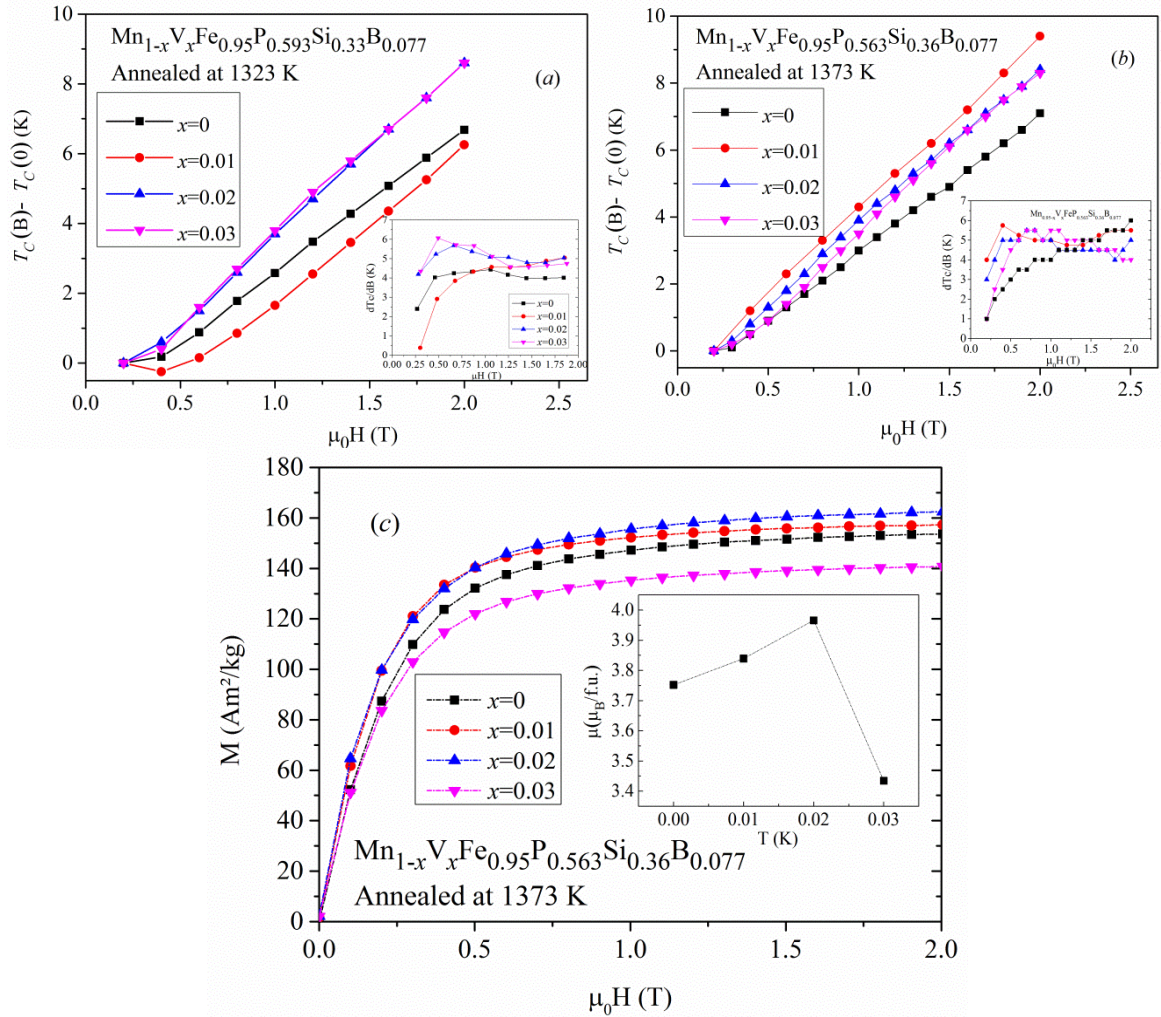

Fig. 4  Field dependence of $T_C$ and $dT_C/dB$ (insets) for *series A* (a) and *series B* (b); (c) The magnetization as a function of the V content for *series A* measured at a temperature of 5 K. The insets are the magnetic moment per formula unit ($\mu_{f.u.}$) dependence of the V content for *series B* .

## 4. Conclusions

The ultra-low hysteresis and giant MCE of $Mn_{1-x}V_xFe_{0.95}P_{0.563}Si_{0.36}B_{0.077}$ alloys annealed at 1373 K paves a path to high frequency magnetic refrigeration applications. $T_C$ tends to decrease with increasing V. For the alloys annealed at 1373 K, the latent heat can be reduced by 55 % from 6.2 to 2.8 J/g and $\Delta T_{hys-MT}$ decreases by 93% from 1.5 to 0.1 K when $x$ increases from 0.00 to 0.03. The field dependence of the transition temperature ($dT_C/dB$) is enhanced from 4.0 to 5.0 K/T by V substitution of Mn. Higher values of $dT_C/dB$ and $\mu_{f.u.}$ value are the key reasons that a large GMCE value can be provided even though hysteresis has been reduced to ultra-low values. Finally, an ultra-low value of $\Delta T_{hys-DSC}$ (0.7 K) and a giant $\Delta T_{ad}$ (2.3 K) can be achieved in a field of 1 T. Thus, the current $Mn_{1-x}V_xFe(P,Si,B)$ compounds can provide a feasible alternative for high-frequency near-room temperature magnetic cooling applications using permanent magnets.

## Acknowledgements

The authors acknowledge Anton Lefering, Kees Goubitz and Bert Zwart for their technical assistance. The authors also thank Yibole Hargen for discussion. This work is part of the Industrial Partnership Program IPP I28 of the Dutch Foundation for Fundamental Research on Matter (FOM), financially supported by BASF New Business. This work is also supported by Guangdong Provincial Science and Technology Program (Grant No. 2015A050502015), the Guangzhou Municipal Science and Technology Program (No. 201505041702137), Natural Science Foundation of the Guangdong Province (2016A030313494), and Zhongshan Municipal Science and Technology Program (Platform construction and innovation team). The author thanks for the financial support of the Guangzhou Ethics Project.